\documentstyle[11pt,newpasp,twoside,epsf]{article}
\font\mitt=cmmi9 scaled\magstep 1

\def\HI{HI}
\def\Halpha{H$\alpha$}
\def\varv{\hbox{\mitt v}}
\def\kms{\ifmmode\mathrm{~km~}\mathrm{s}^{-1}\else km s$^{-1}$\fi}
\def\mlstar{\ifmmode\Upsilon_{\!\!*}\else$\Upsilon_{\!\!*}$\fi}
\def\reference#1{\noindent#1}

\def\edcomment#1{\iffalse\marginpar{\raggedright\sl#1\/}\else\relax\fi}
\reversemarginpar

\begin{document}
\title{The Kinematics in the Cores of Low Surface Brightness Galaxies}
 \author{R.A. Swaters}
\affil{Johns Hopkins University, 3400 N. Charles St., Baltimore MD
21218, U.S.A., and Space Telescope Science Institute, 3700 San Martin
Dr., Baltimore, MD 21218, U.S.A.}
\author{M. A. W. Verheijen}
\affil{Astrophysikalisches Institut Potsdam, An der Sternwarte
16, 14482 Potsdam, Germany.}
\author{M. A. Bershady}
\affil{Astronomy Department, University Wisconsin - Madison, 475 N. Charter St., Madison, WI 53706}
\author{D. R. Andersen}
\affil{Max Planck Institut f\"ur Astronomie, K\"onigstuhl 17,
69117 Heidelberg, Germany}

\begin{abstract}

Systematic effects on HI and H$\alpha$ long-slit observations make a
measurement of the inner slope of the dark matter density distribution
difficult to determine. Halos with constant density cores and ones
with $r^{-1}$ profiles both appear consistent with the data, although
constant density cores generally provide better fits. High-resolution,
two-dimensional velocity fields remove most of the systematic effects,
yet as a result of noncircular and random motions the inner slopes
still cannot be accurately measured.  Halo concentration parameters
provide a more useful test of cosmological models because they are
more tightly constrained by observations.  The concentration
parameters for LSB galaxies appear consistent with, but on the low end
of the distribution predicted by CDM.

\end{abstract}

\section{Introduction}

Studies of dark matter in spiral galaxies have been plagued by
uncertainties in the mass-to-light ratio (M/L) of the stellar disk,
resulting in a large degeneracy in the mass modeling (e.g., van Albada
et al. 1985). This degeneracy can largely be avoided by studying dwarf
and low surface brightness (LSB) galaxies. Even though the central
parts of the observed rotation curves can in principle be explained by
scaling up the contribution of the stellar disks (Swaters 1999;
Swaters, Madore, \& Trewhella 2000), the inferred stellar M/Ls are
much higher than expected from population synthesis modeling (e.g.,
Bell \& de Jong 2001).  For reasonable M/Ls, these galaxies are
dominated by dark matter at all radii, making dwarf and LSB galaxies
ideal for studying the properties of dark matter.

The power law slopes of the central dark matter density distributions
$\rho(r)\propto\rho^{-\alpha}$ are of particular interest.
Cosmological simulations indicate $\alpha$ depends on the nature of
dark matter. A measurement of $\alpha$ may thus provide constraints on
the nature of dark matter and theories of galaxy formation.
Unfortunately, it is difficult to measure the inner slope $\alpha$
accurately, both from simulations (e.g., Power et al. 2003; White,
this volume; Navarro, this volume), and from observations, as is
discussed in detail below (see also de Blok, this volume).

\section{Systematic effects on rotation curves}

The main assumption in using rotation curves to infer the properties
of dark matter halos is that the rotation velocities reflect the
gravitational potential. However, observational effects, such as poor
spatial resolution or slit offsets, and intrinsic properties of the
galaxies, such as pressure support or noncircular motions, may affect
the derived rotation curves and lead to systematic uncertainties in
the derived dark matter properties.

\subsection{HI observations}

Earlier HI studies found that dwarf and LSB galaxies have slowly
rising inner rotation curves (see e.g., Swaters 1999 and references
therein).  Because of their shallow slopes, these rotation curves were
found to be inconsistent with the steep halos predicted by cold dark
matter (e.g., Navarro et al. 1996; McGaugh \& de Blok 1998).  More
recent studies raised the concern that some of these \HI\ rotation
curves may be affected by beam smearing, and found that the
rotation curves of dwarf and LSB galaxies rise more steeply when beam
smearing is taken into account (Swaters 1999; Blais-Ouellette et
al. 1999; Swaters et al.\ 2000).

However, beam smearing in itself does not constitute an unsurmountable
problem. It is possible to correct the measured rotation curves for
the effects of beam smearing (Swaters 1999), or its effects can be
incorporated in the modeling procedure (e.g., van den Bosch et al.\
2000). Nonetheless, because of their relatively low resolution, \HI\
observations may not be best suited to determine the inner rotation
curve slopes. High angular resolution observations, such as \Halpha\
long slit or Fabry-Perot spectroscopy, thus seem more suited.

\subsection{H$\alpha$ long-slit observations}

Despite the high spatial resolution of H$\alpha$ long-slit
observations, it remains difficult to measure $\alpha$
accurately. Different studies find conflicting results: some find LSB
galaxies to be consistent with steep inner slopes (e.g., Swaters et
al. 2003a, hereafter SMvdBB), yet others find them inconsistent with
steep slopes (e.g., de Blok et al. 2001a,b; de Blok \& Bosma 2002;
Marchesini et al.\ 2002). This inconsistency remains even when
identical datasets are used (SMvdBB).

This apparent discrepancy may well be the result of systematic effects
on the long-slit observations, which all lead to an underestimate of
the inner slope. Several effects may contribute: 1) Seeing tends to
wash out the inner gradient of the rotation curve, leading to an
underestimate of the inner slope. 2) If the slit is opened wider,
parts of the galaxy with lower radial velocities will also be visible
through the slit, diluting the inner gradient and resulting in an
underestimate of the inner slope. 3) Slit offsets (the combined
results of telescope pointing errors, errors in the coordinates of the
optical centers, and offsets between optical and dynamical centers)
will cause the steepest gradient in the rotation curve to be missed,
leading to an underestimate of the inner slope. 4) At inclinations
close to edge-on, one line of sight samples a large part of the disk,
resulting in a range of observed velocities. The rotation velocity is
represented by the extreme velocity along the
line-of-sight. Obviously, the usual method of deriving the rotation
velocity by Gaussian fits or a intensity weighted mean will result in
an underestimate of the rotation velocity, which in turn leads to an
underestimate in the inner slope. On top of that, extinction and the
distribution of H$\alpha$ (e.g., a ring) may also lead to an
underestimate of the inner slope. 5) If the H$\alpha$ emission is
concentrated toward the edge of the slit, the derived velocities will
be lower, especially near the center of the galaxy, leading to an
underestimate of the inner slope.  In summary, all systematic effects
lead to an underestimate of the inner slope, so there clearly is an
strong observational bias against finding steep slopes from long-slit
observations.

\section{The inner slope: poor constraints on cosmological models}

In the literature, three principal ways of measuring the inner slope
of the halo density profile have been used: inversion of the rotation
curve (e.g., de Blok et al.\ 2001, SMvdBB), fitting a
power law to the observed rotation curve (e.g., Simon et al.\ 2003),
and fitting a general mass model to the observed rotation curve (e.g.,
van den Bosch et al.\ 2000, SMvdBB).

\subsection{Rotation curve inversion}

If one assumes the disks are dynamically insignificant, and if one
assumes furthermore that the dark matter has a spherically symmetric
distribution, it is possible to recover the density distribution of
the dark matter from the observed rotation curve in a non-parametric
way.  From $\nabla^2\Phi=4\pi G\rho$ and $\Phi=-GM/r$ the density
distribution is given by
\begin{equation}
\rho(r) = {{1}\over{4\pi
    G}}\left(2{{\varv}\over{r}}{{\partial\varv}\over{\partial r}} +
    {{\varv^2}\over{r^2}} \right).
\label{eq1}
\end{equation}
The inner slope $\alpha$ can be measured by inverting the measured
rotation curve using Eq.~\ref{eq1}, and fitting a simple
power-law $\rho \propto r^{-\alpha_m}$ to the density distribution
inside the break radius (the radius where the density profile turns
from its inner slope to its outer $r^{-2}$ slope).

\begin{figure}[t]
\plotfiddle{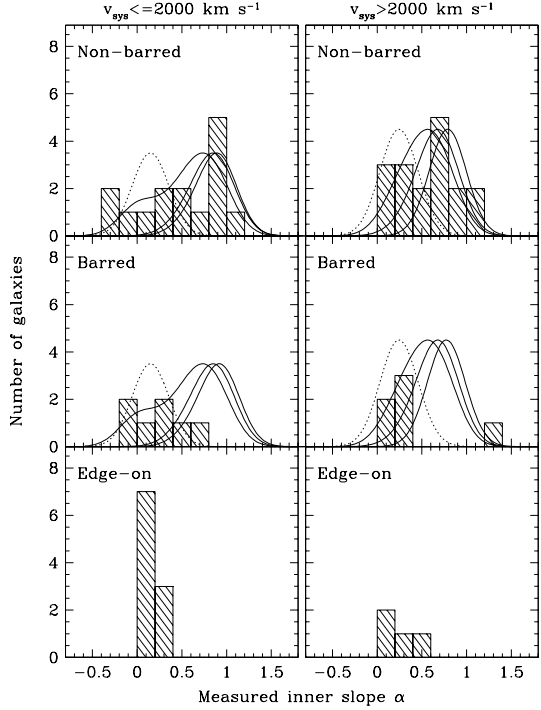}{9cm}{0}{37}{37}{-105}{-20}
\caption{The distribution of measured inner slopes for the galaxies in
the samples of de Blok et al.\ (2001a), de Blok \& Bosma (2002), and
SMvdBB), divided into non-barred, barred, and edge-on
galaxies, both for galaxies within $2000 \kms$ (left), and for
galaxies beyond $2000 \kms$.  Dotted lines represent the expected
distributions of inner slopes for $\alpha=0$, the solid lines from
left to right give the expected distributions for $\alpha=1$, with a
slit alignment error of $2''$, $1''$ and $0''$.\label{fig1}}
\end{figure}

In Fig.~\ref{fig1} we present the distribution of measured inner
slopes for the galaxies in the samples of de Blok et al.\ (2001a), de
Blok \& Bosma (2002), and SMvdBB). Overplotted on the histograms are
the expected distributions of measured inner slopes based on detailed
modeling of the observations, including the effects of distance,
seeing, slit width, and slit alignment errors. See SMvdBB for more
details on the modeling.

For the non-barred galaxies the measured inner slopes span a wide
range from $\alpha=0$ to $\alpha=1$.  The distribution of $\alpha_m$
for the non-barred galaxies in the combined sample appears
inconsistent with the distribution expected for halos with $\alpha=0$.
However, it is possible that the stellar disks do contribute
significantly and this may explain the wing to high $\alpha_m$ even if
the halos have $\alpha=0$.  The models with $\alpha=1$ appear to
underpredict the number of galaxies with $\alpha_m$ near zero, unless
the error in the slit positioning is about $2''$. An error of $2''$ in
the slit positioning may not be unreasonable given the large angular
size of these galaxies and their irregular, diffuse and low surface
brightness nature.

\begin{figure}
\plotone{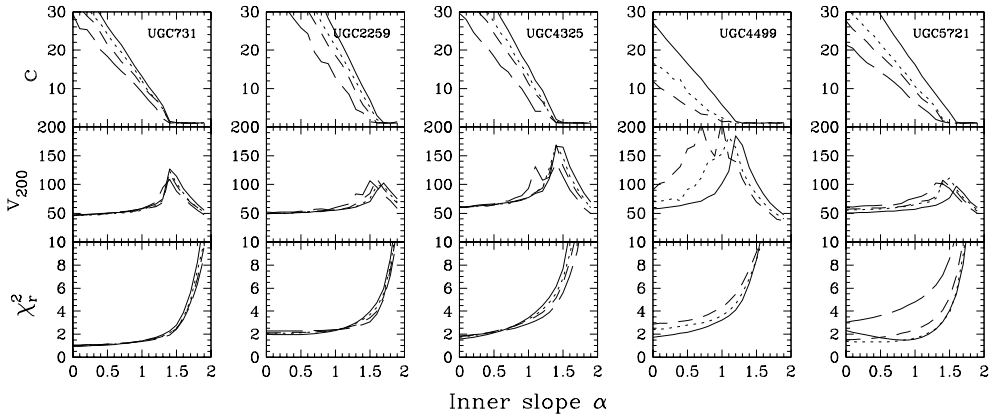}
\caption{From top to bottom the best fit $c$, best fit $\varv_{200}$,
and reduced $\chi^2$for the generic halo fits.  The the solid line
represents $\mlstar=0$ (for the $R$-band), the dotted line is
$\mlstar=0.5$, the short dashed lines is $\mlstar=1$ and the long
dashed line is $\mlstar=2$. Halos with $0<\alpha<1$ are consistent
with most galaxies, although halos with $\alpha=0$ generally provide
better fits.
\label{fig2}}
\end{figure}

Although based on only a few galaxies, the barred galaxies on average
appear to have a somewhat lower $\alpha_m$.  This may be an indication
that the non-circular motions associated with the bar have affected
the inner slopes. Another striking feature in Fig.~\ref{fig1} is the
large difference in the distribution of measured inner slopes between
edge-on galaxies and non-barred galaxies. This difference is most
likely artificial: the method with which the rotation curves were
derived lead to a severe underestimate of the inner slopes (SMvdBB).

Although the inner slope derived from rotation curve inversion has as
advantage that it is a parameter-free measurement and that it directly
measures the inner slope of the halo, it is particularly sensitive to
systematic effects mentioned above because the inner slope is usually
measured from only the innermost points. As a result, it is not
possible to measure the inner slopes accurately, and our conclusion is
that inner slopes in the range $0<\alpha<1$ are consistent with the
observations discussed here.

\subsection{Fitting a power law to the rotation curve}

Some authors have determined the inner slope by fitting a power law
slope to the entire observed rotation curve (e.g., Simon et al.\
2003). Although this gives a more accurate measurement than rotation
curve inversion, it is seems unlikely that an inner slope derived from
a fit to the entire rotation curve can be directly compared to the
numerical simulations, because in the simulations the slopes changes
with radius. A measurement of the true inner slope could be derived by
fitting a power law slope to the inner parts of the rotation curve
only, but that would make the measurement sensitive to the systematic
effects described in the previous subsection.

\begin{figure}
\plotfiddle{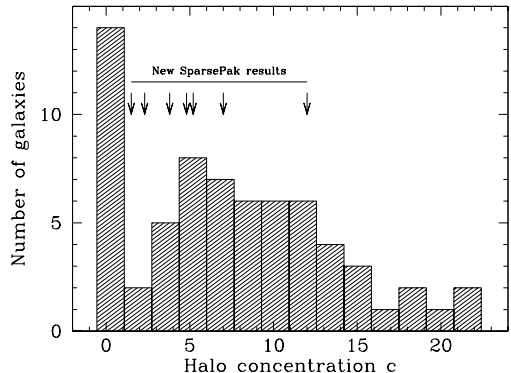}{4.5cm}{0}{25}{25}{-85}{0}
\caption{The shaded histogram gives the distribution of best fit $c$
parameters for galaxies in the samples of de Blok et al.\ (2001a), de
Blok \& Bosma (2002), and SMvdBB\label{fig3}. The arrows indicate the
best fit $c$ parameters derived from new SparsePak data (Swaters et
al.\ 2003, Swaters et al.\ in preparation).}
\end{figure}

\subsection{Fitting a general density profile to the rotation curve}

The inner slope can also be measured by fitting a halo model to the
rotation curve based on a general density profile of the form:
\begin{equation}
\label{eq2}
\rho(r) = {\rho_0 \over (r/r_s)^{\alpha} (1 + r/r_s)^{3-\alpha}},
\end{equation}
where $\rho_0$ is the central dark matter density, and $r_s$ a scale
radius.  This density distribution changes from $\rho \propto
r^{-\alpha}$ for $r \ll r_s$ to $\rho \propto r^{-3}$ for $r \gg
r_s$. For $\alpha=0$ the density profile thus has a constant density
core and becomes comparable to that of the pseudo-isothermal sphere,
while for $\alpha=1$ it reduces to the NFW profile (Navarro et
al. 1997).

The density and scale radius derived from such mass models are less
sensitive to the effects of seeing, slit width and slit offsets,
because the entire rotation curve is used to determine the parameters,
not only the innermost part. Consequently, these mass models are not
very sensitive to the inner slope of the dark matter halo, and for
most galaxies we find that mass models based on $\alpha=1$ provide
fits of similar quality as those based on $\alpha=0$ (see examples in
Fig.~\ref{fig2}). This is not only true if the contribution of the
stellar component is ignored, but also for models in which the stellar
component play a modest role.

From these general mass models we find that for most galaxies the
inner slope is unconstrained, and fits of similar quality are obtained
for the range $0<\alpha<1$, although $\alpha=0$ is somewhat
preferred. Inner slopes as steep as $\alpha=1.5$ are clearly ruled
out.

\begin{figure}
\plotone{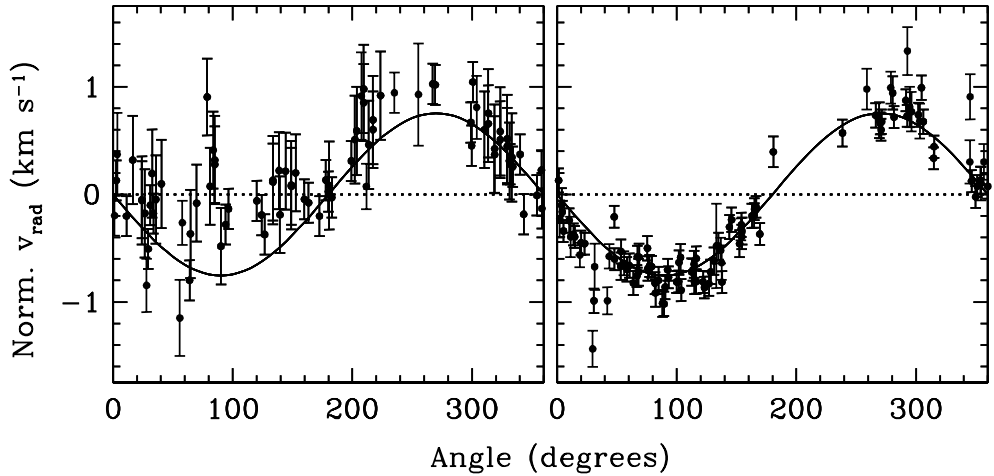}
\caption{Radial velocity relative to the systemic velocity,
normalized by rotation velocity, as function of angle in the
plane, for DDO~39 (obtained with SparsePak on the WIYN telescope, Swaters et al.\ 2003). The left panel shows points within $r<1$~kpc, the right one points
between $1.5<r<2.5$ kpc.  The solid line represents circular
rotation. \label{fig4}}
\end{figure}

\section{Halo concentration: useful constraints on cosmological models}

A different way of testing the predictions of CDM simulations is
through the halo concentration parameter $c$. For an NFW halo
(Eq.~\ref{eq2} with $\alpha=1$), $c=r_{200}/r_s$, where $r_{200}$ is
the virial radius. An important advantage of using the halo
concentration parameter is that it is less sensitive to systematic
effects because it is derived from a fit to the rotation curve as a
whole.  The expected values for $c$ depend on the assumed cosmological
parameters. For a $\Lambda$CDM cosmology, expected values have a
$2\sigma$ range from 5 to 25, with an average of around 10 to 15
(Navarro et al.\ 1997, Bullock et al.\ 2001). In Fig.~\ref{fig3} the
distribution of best fit $c$ values is shown; the uncertainty on $c$
from rotation curve fits is about 20\% to 40\% for most galaxies. The
peak near $c=1$ is likely to be artificial, resulting predominantly
from galaxies in which the fit is unconstrained, such as galaxies with
rotation curves that extend less than two disk scale lengths, and from
edge-on galaxies with possibly incorrect rotation curves.  The rest of
the distribution shows a large spread, and seems somewhat skewed
toward smaller values of $c$ than expected in $\Lambda$CDM.

Low values for $c$ do not necessarily indicate an inconsistency
with $\Lambda$CDM. For example, a bias toward low $c$ values in LSB
galaxies could be explained if LSB galaxies preferentially form in low
density halos. In addition, the concentration parameter also depends
on the slope of the power spectrum of density fluctuations, and $c\sim
5$ is in agreement with models in which structure formation on small
scales is suppressed (Zentner \& Bullock 2002).

\section{High Resolution, Two-Dimensional Velocity Fields}

To avoid systematic effects of long-slit and HI observations, and to
map possible non-circular motions, high spatial resolution,
two-dimensional velocity fields are needed. Several such studies are
now available (e.g., Blais-Ouellette et al.\ 2001; Bolatto et al.\
2002; Swaters et al.\ 2003b; Simon et al.\ 2003).

Swaters et al.\ (2003b) studied the kinematics in the core of the LSB
galaxy DDO~39 with the SparsePak integral field spectrograph on the
WIYN telescope (Bershady et al., submitted), and found random motions
with an average amplitude of about 8 \kms\ throughout the galaxy, and
in addition detected distinct noncircular motions as can be seen in
Fig.~\ref{fig4}. We (Swaters et al. in preparation) have obtained data
for an additional 6 LSB galaxies with SparsePak. Preliminary analysis
of these data again reveal significant random and noncircular motions.
The origin of either component is unclear, but they cause inner slopes
in the range $0<\alpha<1$ to be consistent with the data.  Therefore,
even though these data are a significant improvement on previous \HI\
or \Halpha\ long-slit data, the inner rotation curve, and hence the
inner slope of the dark matter distribution, is still uncertain due to
significant noncircular and random motions.

Fortunately, the availability of two-dimensional velocity fields makes
it possible to determine the kinematic center and estimate the
influence of noncircular motions on the derived rotation curves,
allowing for an accurate measurement of all but the inner rotation
curve. Consequently, halo parameters, such as $c$, that depend on the
entire rotation curve shape can be determined accurately (with a
typical uncertainty on $c$ of about 10\% to 20\%). For DDO~39 and for
the new SparsePak data, we found $c$ values that are on the low side
when compared to simulations (see Fig.~\ref{fig4}).

\section{Conclusions}

The inner slope $\alpha$ is difficult to measure from \HI\ and
H$\alpha$ long-slit observations because of systematic
effects. High-resolution, two-dimensional velocity fields provide a
major step forward, but deviations from circular motions still make an
accurate measurement of $\alpha$ difficult. A range of inner slopes
$0<\alpha<1$ appear consistent with the data, although models with
$\alpha=0$ generally provide somewhat better fits. Thus, it seems that
an accurate measurement of $\alpha$ and, from that, a tight limit on
the nature of dark matter may be difficult to achieve observationally
from H$\alpha$ emission, which is susceptible to kinematic
perturbations.

Halo parameters that have been derived from the entire rotation curve
instead of only the central parts, such as the halo concentration
parameter $c$, may provide a more robust test for cosmological
models. In the case of NFW halos, the halo concentration parameters of
LSB galaxies tend to be lower than expected from cosmological
simulations.

\bigskip{\bf \noindent References}

{\small 

\reference{}Bell, E.\ F.\ \& de Jong, R.\ S.\ 2001, \apj, 550, 212

\reference{}Blears-Ouellette, S.\, Carignan, C., Amram, P., \& C{\^
  o}t{\' e}, S.\ 1999, \aj, 118, 2123

\reference{}de Blok, W. J. G., \& Bosma, A. 2002, \aap, 385, 816

\reference{}de Blok, W.\ J.\ G., McGaugh, S.\ S., Bosma, A., \&
  Rubin, V.\ C.\ 2001, \apjl, 552, L23

\reference{}Bullock, J.~S., Kolatt, T.~S., Sigad, Y., Somerville,
  R.~S., Kravtsov, A.~V., Klypin,\break\leavevmode\null\kern0.67cm A.~A., Primack, J.~R., \& Dekel, A.\
  2001, \mnras, 321, 559

\reference{}McGaugh, S. S., de Blok, W. J. G. 1998, \apj, 499, 41

\reference{}Navarro, J.\ F., Frenk, C.\ S., \& White, S.\ D.\ M.\ 
  1996, \apj, 462, 563

\reference{}Navarro, J.\ F., Frenk, C.\ S., \& White, S.\ D.\ M.\ 
  1997, \apj, 490, 493

\reference{}Power, C., et al., 2003, MNRAS, 338, 14

\reference{}Simon, J.~D., Bolatto, A.~D., Leroy, A., \& Blitz, L.\
  2003, \apj, 596, 957

\reference{}Swaters, R. A. 1999, PhD thesis, Rijksuniversiteit
  Groningen

\reference{}Swaters, R.\ A., Madore, B.\ F., \& Trewhella, M.\ 
  2000, \apjl, 531, L107

\reference{}Swaters, R. A., Madore, B. F., van den Bosch, F.,
  Balcells, M., \apj, 583, 732 (SMvdBB)

\reference{}van Albada, T. S., Bahcall, J. N., Begeman, K., Sancisi,
  R. 1985, \apj, 295, 305

\reference{}van den Bosch, F.\ C., Swaters, R. A. 2001, \mnras, 325,
1017

\reference{}van den Bosch, F.\ C., Robertson, B.\ E., Dalcanton,
  J.\ J., \& de Blok, W.\ J.\ G.\ 2000,\break\leavevmode\null\kern0.67cm\aj, 119, 1579

\reference{}Zentner, A.~R.~\& Bullock, J.~S.\ 2002, \prd, 66, 43003

\end{document}